\documentclass[12pt]{article}
\usepackage{amssymb}
\usepackage{epsfig}
\usepackage{float}

\newcommand{\be}{\begin{equation}}
\newcommand{\ee}{\end{equation}}
\newcommand{\bea}{\begin{eqnarray}}
\newcommand{\eea}{\end{eqnarray}}

\begin{document}
\begin{flushright}
{hep-ph/0512332}
\end{flushright}

\vskip 1cm

\begin{center}
{\bf TOWARDS A GEOMETRIC APPROACH TO THE FORMULATION OF THE STANDARD
MODEL}
\end{center}

\vskip 1cm
\begin{center}
V.G. Kadyshevsky$^\ast$, M.D. Mateev$^\ast$$^\dag$,
V.N. Rodionov$^\ddag$, A.S. Sorin$^\ast$
\end{center}
\vspace{0.1cm}
\begin{center}
{$^\ast$\em  Joint Institute for Nuclear Research,\\ 141980 Dubna (Moscow Region), Russia\\
kadyshev@jinr.ru \quad sorin@theor.jinr.ru}
\end{center}
\begin{center}
{\dag\em University of Sofia St. Kliment Ohridsky, Sofia, Bulgaria\\ matey.mateev@gmail.com}
\end{center}

\begin{center}
{\ddag\em Moscow State Geological Prospecting University, 118617 Moscow,
Russia\\ vnrodionov@mtu-net.ru}
\end{center}

\vskip 1cm {\bf Abstract.}{
A geometric interpretation of the spontaneous symmetry breaking effect,
which plays a key role in the Standard Model, is developed. The advocated
approach is related to the effective use of the momentum 4-spaces of
the constant curvature, de Sitter and anti de Sitter, in the apparatus
of quantum field theory.}

\vskip 1cm

\section{Introduction}

We sent this work for publication at the end of 2005 marked as the
\emph{\textbf{World Year of Physics}}. As its epigraph we could use the
well-known citation of A. Einstein:
\begin{equation}\label{4}
     \emph{\rm{EXPERIMENT = GEOMETRY + PHYSICS}}\,
\end{equation}
This thesis has been convincingly confirmed in the special theory of
relativity, the general theory of relativity and quantum theory. The universal
constants $c$, $G$ and $\hbar$ playing a key role in these theories, allow
simple interpretation in either a geometry context or group theoretical terms
directly connected with geometry.

In the special theory of relativity the velocity of light $c$ appears in
the definition of the metric of the pseudo-Euclidean space-time:
\begin{equation}\label{1}
ds^2 = c^2 dt^2 - (dx_1)^2-(dx_2)^2-(dx_3)^2.
\end{equation}
The corresponding 3-dimensional velocity space has Lobachevsky geometry with
negative curvature $-1/c^2$.

In the general theory of relativity the space time carries Riemann
geometry and the components of the Riemann metric tensor depending  on the
universal constants $c$ and $G$ as parameters play the role of
gravitational potentials.

In quantum mechanics such observables as momentum, energy, and angular momentum
are generators of the Galilean group. The Planck constant $\hbar$ is involved
in them explicitly  and determines the value of the quantum of angular
momentum.

Taking into account the above considerations we may formulate the
\emph{\textbf{heuristic conclusion:}} the appearance of  universal
constants in the description of certain physical phenomena is a signal
that in such cases an important role may be played by
\emph{\textbf{geometric}} arguments which, because of (\ref{4}), simplify
the understanding of the \emph{\textbf{physical }}essence of the
considered processes.

To this end  let us turn to the contemporary theory of elementary particles
known as the Standard Model (SM). The notion "elementary particle"
supposes that in accordance with present experimental data these objects do
not have a composite structure and are described by \emph{\textbf{local
fields}}. The SM Lagrangian depends on a \emph{\textbf{finite}} number of
fields of this kind:

- three families of quark and lepton fields;

- four vector boson fields $W^{\pm}$, $Z^0$, $\gamma$;

- an octet of gluon fields $g$;

- the hypothetic field of the Higgs boson $H$.

The $H$-boson has a different from zero vacuum expectation value
\begin{equation}\label{2}
    \langle0|H(x)|0\rangle = h_0
\end{equation}
with dimension of mass. Thanks to this circumstance  fields which have to
be massive obtain masses proportional to $h_0$ when interacting with
$H(x)$. As far as in the SM there are a finite number of particles, their
masses are bounded from above.

The main idea underlying this work is a more general statement:
\emph{\textbf{mass spectrum of all particles described by local fields
should cuts off on a certain mass $M$:}}
\begin{equation}\label{3}
    m \leq M.
\end{equation}
This universal parameter is called the \emph{\textbf{fundamental mass}}.
In other words, in fact we introduce \emph{\textbf{a new notion of a local
field}}, intrinsically consistent with the condition (\ref{3}). Now
objects for which $ m
> M$ cannot be considered as elementary particles, as  to them does not
correspond a local field.

The above-presented approach allows a simple geometric realization if one
considers that the fundamental mass $M$ is the curvature radius of the momentum
anti de Sitter 4-space ($\hbar = c =1$)
\begin{equation}\label{O32}
    p_0^2 - p_1^2 - p_2^2 - p_3^2 + p_5 ^2 = M^2.
\end{equation}
For a free particle, for which  $p_0^2 - \overrightarrow{p}^2 = m^2$, the
condition (\ref{3})  is automatically satisfied on the surface
(\ref{O32}). In the approximation
\begin{equation}\label{Plpred}
    |p_0|,\;\;|\overrightarrow{p}| \ll M, p_5 \cong M.
\end{equation}

\noindent the anti de Sitter geometry does not differ from the Minkowski
geometry in four dimensional pseudo--Euclidean $p$-space.

However, it is much less obvious that in  the momentum 4-space (\ref{O32})
one may fully develop the apparatus of quantum field theory, which after
transition to configuration representation (with the help of a specific
5-dimensional Fourier transform) looks like a local field theoretical
formalism in the four dimensional $x$-space \cite{Kad1}--\cite{Kad2}.
It is fundamentally important that the new theory may be formulated in a
gauge invariant way \cite{Kad1,KM1,CDKM}). In other
words, in the considered \emph{\textbf{geometric}} approach there are
conditions to construct an adequate generalization of the Standard Model,
which gives one more illustration of the profundity of the Einstein idea
(\ref{4}). The following sections of this paper are devoted to this task.

At the end of this introduction we would like to note  one important
geometric point. The fact is that simultaneously with (\ref{O32}) , in
$5$-dimensional $p$-space a second surface exists,
\begin{equation}\label{O41}
    p_0^2 - p_1^2 - p_2^2 - p_3^2 - p_5 ^2 = - \mathfrak{M}^2 ,
\end{equation}
on which the  \emph{\textbf{$4$-dimensional de Sitter  momentum space }}is
realized. Although in this case the boundary on the particles mass
(\ref{3}) does not appear, in the flat limit (\ref{Plpred}) the de Sitter
geometry, similarly to anti de Sitter one, coincides with the
pseudo-Euclidean geometry of Minkowski $p$-space. Moreover, on the basis
of (\ref{O41}) it is not difficult to develop the formalism of local
gauge-invariant quantum field theory in configurational 4-space
\cite{Kad1}--\cite{Kad2}. In what follows we shall see that this
version of the theory will find its application in the realization of the
Higgs mechanism in the developed approach.

\section{Spontaneous symmetry breaking in the case of neutral self--interacting scalar
field}

Let us demonstrate that in the theory of self--interacting neutral scalar
field $\varphi(x)$ developed on the basis of the de Sitter geometry
(\ref{O41}) there is a spontaneous symmetry breaking of  discrete symmetry
(an analogue of $\varphi(x) \rightarrow - \varphi(x)$). Moreover, after
the transition to the new stable vacuum the Lagrangian of the system takes
the form corresponding to the anti de Sitter case (\ref{O32}) and the
fundamental mass $M$ is determined in terms of $\mathfrak{M}$. In other
words, there is a phase transition from the de Sitter geometry to anti
de Sitter geometry.

In accordance to \cite{KM2}, the free field Lagrangian for the massless
 neutral scalar field in the case (\ref{O41}) has the
form:
\begin{equation}\label{LO41m0}
    \mathcal{L}_0(x) = \frac{1}{2}\left(\frac{\partial\varphi(x)}{\partial
    x_\mu}\right)^2 + \frac{\mathfrak{M}^2}{2}(\chi(x) -
     \varphi(x))^2 \equiv \frac{1}{2}\left(\frac{\partial\varphi(x)}{\partial
    x_\mu}\right)^2 - U_0 (\varphi, x),
\end{equation}

\noindent where $\chi(x)$ is a neutral auxiliary field.

If we now introduce in (\ref{LO41m0}) the simplest interaction in the
 form\footnote{Let us note \cite{KM2} that in the limit
$\mathfrak{M}\rightarrow\infty$, i.e. after the transition to the theory
in the Minkowski $p$-space, the field variables $\varphi(x)$ and $\chi(x)$
coincide: $\varphi(x)=\chi(x)$.}:

\begin{equation}\label{L41int}
   \mathcal{L}_{int}(x) = - \frac{\lambda^2}{4}\left(\varphi^2(x) + \chi^2(x)\right)^2,
\end{equation}

\noindent we obtain the following expression for the total density of the
potential energy:
\begin{equation}\label{Ux}
    U(x) = - \frac{\mathfrak{M}^2}{2}\left(\chi(x) -
     \varphi(x)\right)^2 + \frac{\lambda^2}{4}\left(\varphi^2(x) + \chi^2(x)\right)^2.
\end{equation}

\noindent It is evident that the considered system has a nonstable vacuum
state. The spontaneous transition to the new stable vacuum is accompanied by
breaking of the discrete symmetry:
\begin{equation}\label{dissymm}
    \begin{array}{c}
      \varphi(x) \rightarrow - \varphi(x)\\
 \chi(x)  \rightarrow   - \chi(x). \\
    \end{array}
\end{equation}
The standard procedure of substitution of $\varphi(x)$ and $\chi(x)$ by
variables with zero vacuum expectation values

\begin{equation}\label{phichiprime}
    \varphi'(x)= \varphi(x)- \varphi_0 = \varphi(x)-
    \frac{\mathfrak{M}}{\lambda},\;\;\;
    \chi'(x)= \chi(x) - \chi_0 = \chi(x) + \frac{\mathfrak{M}}{\lambda}
\end{equation}
leads to the total Lagragian of the system of the form:
\begin{equation}\label{L0ev}
\begin{array}{c}
\mathcal{L}(x) = \frac{1}{2}\left(\frac{\partial\varphi'(x)}{\partial
    x_\mu}\right)^2 - \frac{\mathfrak{M}^2}{2}(\chi'(x) -
     \varphi'(x)+ 2\frac{\mathfrak{M}}{\lambda})^2 + \\
    + \frac{\lambda^2}{4}\left((\varphi'(x) - 2\frac{\mathfrak{M}}{\lambda})^2 +
     (\chi'(x)+2\frac{\mathfrak{M}}{\lambda})^2\right)^2 =   \\
= \frac{1}{2}\left(\frac{\partial\varphi'(x)}{\partial
    x_\mu}\right)^2 - \frac{3}{2}\mathfrak{M}^2(\varphi'^2(x) + \chi'^2(x))
    + \mathfrak{M}^2\varphi'(x) \chi'(x) + \mathcal{L}_{int}= \\
    = \mathcal{L}_0(\varphi', \chi') + \mathcal{L}_{int}(\varphi', \chi'). \\
    \label{lll}
\end{array}
\end{equation}
On the other hand, in the theory of the neutral scalar field with mass
$m$, based on the anti de Sitter geometry (\ref{O32}), the free field
Lagrangian $\mathcal{L}_0(\varphi, \chi)$  has the form \cite{KM2,Kad2}:
\begin{equation}\label{L032}
\begin{array}{c}
    \mathcal{L}(x) = \frac{1}{2}\left(\frac{\partial\varphi(x)}{\partial
    x_\mu}\right)^2 - \frac{m^2}{2}\varphi(x)^2 - \frac{M^2}{2}(\chi(x) -
    cos\mu \;\varphi(x))^2 = \\
 = \frac{1}{2}\left(\frac{\partial\varphi(x)}{\partial
    x_\mu}\right)^2 - \frac{M^2}{2}(\varphi^2(x) + \chi^2(x))+ M^2\cos\mu\;\varphi(x)\chi(x), \\
\cos\mu = \sqrt{1 - \frac{m^2}{M^2}}.\\
\end{array}
\end{equation}

Comparing $\mathcal{L}(x)$ (\ref{L032}) and  $\mathcal{L}_0(\varphi', \chi')$ (\ref{lll}), we conclude
that these two expressions are identical if one puts
\begin{equation}\label{mhiggs}
    \mathfrak{M}^2 = M^2 cos\mu ; \;\;\;\;\;\;\; 3\; \mathfrak{M}^2 = M^2.
\end{equation}
Therefore, as a result of the spontaneous symmetry breaking the field
$\varphi(x)$ obtains the mass
\begin{equation}\label{mh}
    m_H = \frac{2\sqrt{2}}{3}M < M.
\end{equation}

 At the end of this paragraph we shall make two remarks:

1. In our approach all fields (including $\varphi(x)$ and, in a more
general case, the Higgs field $H(x)$) before the symmetry breaking may be
considered as massless. The point is that here the dimension of mass $[m]$
is generated by the fundamental mass $\mathfrak{M}$.

2. In contrast to the standard approach we did not need to introduce a
tachyon. In certain sense the role of a tachyon mass is played by the
quantity $\mathfrak{M}$, which is the curvature radius of the de Sitter
$p$-space.

\section{Spontaneous symmetry breaking in the case of interacting abelian vector
and charged scalar fields}

Let us apply the developed in \cite{Kad1}--\cite{Kad2} methods to
describe the interaction between a neutral abelian vector field and a
 charged scalar field with self--interaction. As before we shall start with
 the de Sitter geometry (\ref{O41}). The total Lagrangian of the
 considered system in this case has the form\footnote{With
 the details one may be acquainted, for instance, in \cite{CDKM}.}
 \begin{equation}\label{Ltot}
    \begin{array}{c}
 \mathcal{L}(x) = \Bigg[- \frac{1}{4}\overline{F_{KL}}(x, x_5)F^{KL}(x, x_5) + \\
 + 2\left| \frac{\partial(e^{-i\mathfrak{M} x_5}A_L(x, x_5))}{\partial x_L}
  - 2i \mathfrak{M} e^{-i\mathfrak{M} x_5} A_5(x, x_5)\right|^2      \\
      +
\overline{D_\mu \varphi(x, x_5)}D^\mu \varphi(x, x_5) + \mathfrak{M}^2
|\varphi(x, x_5) - \frac{i}{\mathfrak{M}}D_5\varphi(x, x_5)|^2 -\\
- \lambda^2\left(|\varphi(x, x_5)|^2 +
\frac{1}{\mathfrak{M}^2}|D_5\varphi (x, x_5)|^2 \right)^2\Bigg]_{x_5 = 0}, \quad
K,L = 0, 1, 2, 3, 5
    \end{array}
\end{equation}
where we have introduced the covariant derivatives
\begin{equation}\label{D}
    \begin{array}{c}
     D_\mu = \partial_\mu + i q e^{-i\mathfrak{M}x_5}A_\mu(x, x_5),  \\
      D_5 = \partial_5 + i q e^{-i\mathfrak{M}x_5}A_5(x, x_5). \\
    \end{array}
\end{equation}
Let us emphasize that in (\ref{Ltot}) the expression in square brackets is
defined in five dimensional configuration space. The Lagrangian involves
the variables $\frac{i}{\mathfrak{M}}\frac{\partial A_\mu}{\partial x_5}$
and $\frac{i}{\mathfrak{M}}\frac{\partial A_5}{\partial x_5}$, which have
an auxiliary character. Moreover, (\ref{Ltot})  depends on the component
$A_5$ which is a gauge degree of freedom. One may easily exclude all these
quantities. As a result the Lagrangian takes the form
\begin{equation}\label{Lint41}\begin{array}{c}
  \mathcal{L}(x) = - \frac{1}{4}F_{\mu\nu}(x)F^{\mu\nu}(x) + \overline{D_\mu
\varphi(x)}D^\mu \varphi(x) +                                \\
+ \;\mathfrak{M}^2 |\varphi(x) - \chi(x)|^2 - \lambda^2(|\varphi(x)|^2 +
|\chi(x)|^2 )^2,                                 \\
\end{array}
\end{equation}
where
\begin{equation}\label{chi}
    \chi(x) = \frac{i}{\mathfrak{M}}\frac{\partial\varphi(x, x_5)}{\partial
    x_5}\Bigg|_{x_5= 0}.
\end{equation}
Let us separate the real and imaginary parts in $\varphi$ and $\chi$,
\begin{equation}\label{phi12chi12}
    \varphi = \frac{1}{\sqrt{2}}(\varphi_1 + i\varphi_2),\;\;\;\;\;\;\;\;
    \chi= \frac{1}{\sqrt{2}}(\chi_1 + i\chi_2),
\end{equation}
and pass to fields with zero vacuum expectation values, choosing the
phases in a simple manner (compare with (\ref{phichiprime})).
\begin{equation}\label{Fo}
 \begin{array}{c}
 \varphi_1 \rightarrow \varphi_1 -
\frac{\mathfrak{M}}{\lambda},\;\;\;\;\;\;\;\;\;\chi_1 \rightarrow \chi_1 +
\frac{\mathfrak{M}}{\lambda},                            \\
\varphi_2 \rightarrow \varphi_2,\;\;\;\;\;\;\;\;\;\;\;\;\chi_2\rightarrow
\chi_2. \\
\end{array}
\end{equation}
In the same way as in the usual theory one of the results of the
spontaneous breaking of the gauge symmetry is a rearrangement of the field
degrees of freedom with conservation of their number. The field
$\varphi_2$ becomes the longitudinal component of the vector field $A_\mu$
and the latter obtains the mass
\begin{equation}\label{m2}
    m_V^2 = \frac{\mathfrak{M}^2 q^2}{\lambda^2}.
\end{equation}
Moreover, the quadratic in the scalar fields part of the Lagrangian has a
structure corresponding to anti de Sitter geometry (\ref{O41}).
Therefore, the parameter  $\mathfrak{M}$ may be expressed, using
(\ref{mhiggs}), in terms of the fundamental mass $M$. As a result we
obtain from (\ref{m2}) for the mass $m_V$ of the vector field
\begin{equation}\label{mV}
    \frac{m_V^2}{M^2} = \frac{q^2}{3\lambda^2}.
\end{equation}
Since in the geometry (\ref{O32}) the mass of a particle cannot exceed the
fundamental mass $M$, from (\ref{mV}) it follows the relation between the
coupling constants $q$ and $\lambda$
\begin{equation}\label{qlambda}
    q \leq \sqrt{3}\lambda.
\end{equation}

In the next paper we shall consider the mechanism of spontaneous symmetry
breaking of nonabelean gauge symmetry in the framework of the present
approach.

~{}

~{}

{\bf Acknowledgements}
The authors would like to express  their sincere gratitude to I. Aref'eva
and I. Volovich for fruitful discussions of different aspects of this
work.

The study was partly supported by the Russian Foundation for Basic Research
(project N05-02-16535) and the Council of the President of the Russian Federation
for support of Young Russian Scientists and Leading Scientific Schools
(project NNSh-2027.2003.2)

\end{document}